\documentclass[prb,twocolumn,superscriptaddress]{revtex4}
\usepackage{graphicx,amsmath,amssymb,bm,placeins}
\usepackage{hyperref}
\newcommand{\bsigma}{{\boldsymbol\sigma}}

\begin{document}
\title{Marginality of bulk-edge correspondence for single-valley Hamiltonians}
\author{Jian Li}
\affiliation{D\'{e}partement de Physique Th\'{e}orique,
Universit\'{e} de Gen\`{e}ve, CH-1211 Gen\`{e}ve 4, Switzerland}
\author{Alberto F. Morpurgo}
\affiliation{DPMC and GAP, Universit\'{e} de Gen\`{e}ve, CH-1211
Gen\`{e}ve 4, Switzerland}
\author{Markus B\"{u}ttiker}
\affiliation{D\'{e}partement de Physique Th\'{e}orique,
Universit\'{e} de Gen\`{e}ve, CH-1211 Gen\`{e}ve 4, Switzerland}
\author{Ivar Martin}
\affiliation{Theoretical Division, Los Alamos National Laboratory,
Los Alamos, New Mexico 87545, USA}
\date{\today}

\begin{abstract}
We study the correspondence between the non-trivial
topological properties associated with the individual
valleys of gapped bilayer graphene (BLG), as a prototypical
multi-valley system, and the gapless modes at its edges and
other interfaces. We find that the exact connection between
the valley-specific Hall conductivity and the number of
gapless edge modes does not hold in general, but is
dependent on the boundary conditions, even in the absence of
intervalley coupling. This non-universality is attributed to
the absence of a well-defined topological invariant within a
given valley of BLG; yet, a more general topological
invariant may be defined in certain cases, which explains
the distinction between the BLG-vacuum and BLG-BLG
interfaces.
\end{abstract}
\maketitle

\section{Introduction}

The analysis of the electronic properties of semiconductors
often requires considering the presence in their band
structure of multiple degenerate valleys, centered around
different, symmetry-related positions in reciprocal space
\cite{seeger_semiconductor_1991}. At low energy, the
dynamics of electrons in each of these valleys can be
modeled in terms of a long-wavelength effective Hamiltonian,
an approach suitable to describe the vast majority of the
transport phenomena that are measured experimentally. One
interesting question -- which is particularly timely in view
of the intense research effort focusing on so-called
topological insulators \cite{hasan_topological_2010,
qi_quantum_2010} -- is whether there exist non-trivial
topological properties associated to the individual valleys
that can be described in the framework of a long-wavelength
effective Hamiltonian.

\begin{figure}
\begin{center}
 \includegraphics[width=0.35\textwidth]{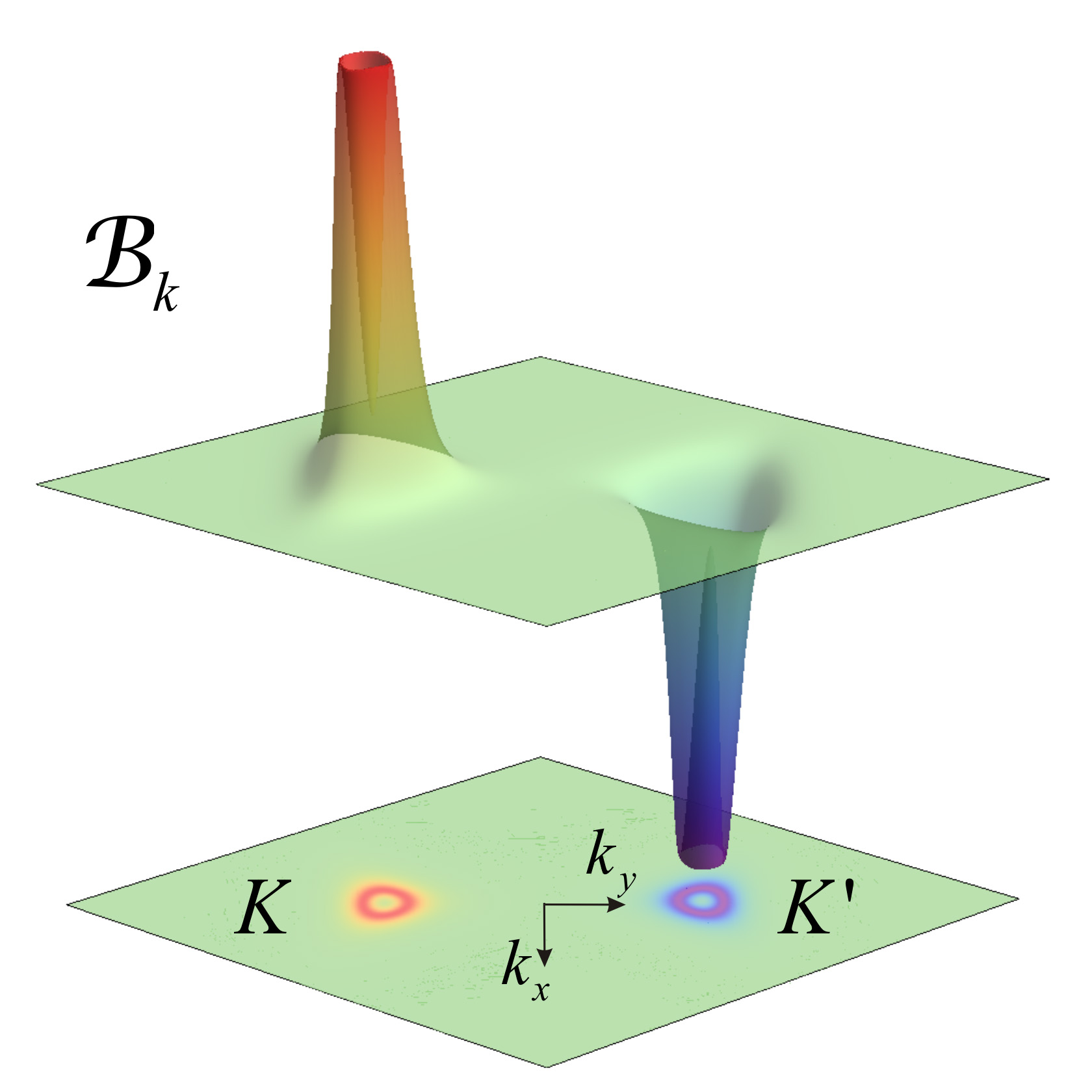}
\end{center}
\caption{Plot of ${\mathcal B}_k = \frac{1}{4\pi} \hat{\bf
g}\cdot [\partial_{k_x} \hat{\bf g}\times\partial_{k_y}
\hat{\bf g}]$ (integrand of Eq. \eqref{eq:intro2}) as a
function of momentum $\bm{k}$ over the entire Brillouin
zone, obtained from the tight-binding description of gapped
BLG. ${\mathcal B}_k$ is nonvanishing only close to the $K$
and $K'$ points (valleys), where it is well approximated by
the expression for $\hat{\bf g}(\bm{k})$ that enters the
long-wavelength effective Hamiltonian (see Eq.
\eqref{eq:ham2}). The integral of ${\mathcal B}_k$ over the
entire Brillouin zone vanishes, owing to the equal and
opposite contributions of the two valleys ($= \pm 1$,
corresponding to the quantized Hall conductivity of the
individual valleys).} \label{fig:tcd}
\end{figure}

To start addressing this question, here we focus on the case
of two-dimensional (2D) electronic systems. The two best
known examples of non-trivial topological insulators in two
dimensions are provided by integer quantum Hall systems
\cite{klitzing_new_1980} and by spin-orbit induced
topological insulators \cite{kane_quantum_2005,
bernevig_quantum_2006}. In these systems, the non-trivial
topological properties of the bulk band structure result in
a quantized Hall conductivity when the Fermi level is
located in a bulk energy gap (for spin-orbit induced
topological insulators, spin-Hall conductivity is quantized
only when the spin is a good quantum number and one can
consider the Hall conductivity for each spin state
separately). Indeed, it is well established that the
expression of the bulk Hall conductivity $\sigma_H$ given by
the Kubo formula for linear response corresponds to the
Chern number that characterizes the topological structure of
the mapping from the Brillouin zone to the space of the
Bloch states \cite{thouless_quantized_1982,
kohmoto_topological_1985}. If we confine ourselves to the
simplest case where only two bands are relevant, the physics
of these systems can be described in terms of an effective
Hamiltonian of the form
\begin{align}\label{eq:intro1}
H=\sum_{\bm{k}}\Psi_{\bm{k}}^\dag[-{\bf g}(\bm{k})\cdot
{\bsigma} + \epsilon_{\bm{k}} - \mu]\Psi_{\bm{k}}
\end{align}
where $\Psi_{\bm{k}} = [c_{\bm{k}\uparrow},
c_{\bm{k}\downarrow}]^T$ is the itinerant electron field
operator,  $\bsigma = (\sigma_x, \sigma_y, \sigma_z)$ is a
vector of the Pauli matrices, and ${\bf g}({\bm{k}})$ is a
real vector. If the bulk band structure is fully gapped, and
the chemical potential $\mu$ lies inside the gap, the Hall
conductivity (in units of $e^2/h$ neglecting spin
degeneracy), equal to the Chern number associated to the
occupied band, is given by \cite{volovik_universe_2003}
\begin{align}\label{eq:intro2}
\sigma_H = N = \frac{1}{4\pi}\int {d^2k} \:\hat{\bf g} \cdot
\left(\partial_{k_x} \hat{\bf g} \times
\partial_{k_y} \hat{\bf g} \right)
\end{align}
with $\hat{\bf g}= {\bf g}/|{\bf g}|$ (again, for spin-orbit
induced topological insulators with conserved spin, these
expressions hold separately for the two spin directions).
The integral has to be performed over the entire Brillouin
zone. The system is topologically non-trivial if  $N \neq
0$. The physical manifestation of the topological
non-triviality, which discriminates between topological and
trivial insulators, is the appearance of $N$ gapless states
at the system edges, which can transport current even when
the Fermi level is located in the bulk energy gap. This
statement, true for both integer quantum Hall and spin-orbit
induced topological insulators with conserved spin, relates
the bulk electronic structure (which determines the Chern
number) to the edge properties and is known as {\it
bulk-edge correspondence} \cite{hatsugai_chern_1993,
qi_general_2006}.

The integrand in Eq.~(\ref{eq:intro2}) is often sharply
peaked in the regions of the momentum space where $|{\bf
g(\bm{k})}|$, and hence the gap, becomes small -- these are
the valleys of the band structure. Thanks to the fast
convergence, the integral in Eq.~(\ref{eq:intro2}) over the
Brillouin zone can be calculated using the long wavelength
approximation for the Hamiltonian valid in the individual
valleys. In this case one can associate an index $N_{\tau}$
to each one of the valleys ($\tau$ labels the valley), and
the Chern number is then the sum of the $N_{\tau}$. This sum
can vanish even if the individual $N_{\tau}$ do not, in
which case the system is --according to the definition given
above-- topologically trivial. However, the question arises
as to whether the non-vanishing of the $N_{\tau}$ associated
to the individual valleys has observable consequences.
Indeed, a non-vanishing $N_{\tau}$ implies that valley
$\tau$ gives an non-zero contribution to the Hall
conductivity, and the appearance of $N_{\tau}$ gapless edge
states associated to each individual valley may be expected
by a naive ``extension" of the bulk-edge correspondence. It
may be argued that due to the overall triviality of the
system (i.e., the fact that the Chern number defined over
the entire Brillouin zone vanishes) these states localize
because they are not protected against disorder at the edge,
which causes inter-valley scattering
\cite{li_topological_2010}. Nevertheless, the fundamental
question remains, whether in a system with ideal edges
(i.e., edges that preserve the {\em valley quantum number},
well defined at low energies) there exists a bulk-edge
correspondence for individual valleys.

Here we consider the case of gapped bilayer graphene (BLG)
as a model system for the case in which two symmetry-related
valleys are present (the $K$ and $K'$ valleys
with non-vanishing $N_{\tau}=\pm 1$ depending on the valley, see Fig.
\ref{fig:tcd}), and investigate the low-energy
electronic states at its edges. Specifically, we consider
different crystalline edges which do not couple the valleys,
and we find the corresponding edge states in the different
cases (we have discussed elsewhere the effect of disordered
edges, which is experimentally more
relevant\cite{li_topological_2010}). If the bulk-edge
correspondence could be generically extended to
single-valley Hamiltonians, one would expect that
$N_{\tau}=\pm 1$ should imply the presence of exactly one
gapless mode per valley per spin at an ideal edge of gapped
BLG, for all edges that do not couple the valleys. In
contrast to this expectation, we show through an explicit
analytical solution in complete agreement with
full tight-binding calculations, that the number of the
gapless edge modes depends on the boundary conditions, even
when no mixing between valleys exists at the
edge\cite{footnote1}. In other words, {\em the
valley-specific bulk-edge correspondence is not fulfilled
for plain BLG-vacuum interfaces.}

In order to understand this result, we examine the
geometrical meaning of $N_{\tau}$ associated to a particular
valley Hamiltonian, and find that, contrary to the case of
the Chern numbers in topological insulators (defined over
the entire Brillouin zone), $N_{\tau}$ in BLG does not
correspond in itself to a well-defined topological invariant
of a mapping. Nevertheless, the non-vanishing of $N_{\tau}$
still signals that the properties of the electron states in
individual valleys are non-trivial, and we discuss how this
non-triviality manifests itself at interfaces between
different domains where the gap of BLG changes sign
\cite{martin_topological_2008}. In this case, the number of
zero-energy states corresponds to the difference of
$N_{\tau}$ on opposite sides of the BLG-BLG interface
($N_{\tau}^l-N_{\tau}^r$, where we denote with
$N_{\tau}^{l,r}$ the value of $N_{\tau}$ at the left and at
the right of the BLG-BLG interface), in agreement with
expectations based on bulk-edge correspondence. Indeed, we
show that for a BLG-BLG interface the difference
$N_{\tau}^l-N_{\tau}^r$ is a well-defined topological
invariant of a mapping, even though the individual
$N_{\tau}^l$ and $N_{\tau}^r$ are not. These results reveal
the {\em marginal} character of individual valleys in gapped
BLG: the {\em difference} of $N_{\tau}$ across an interface
may or may not be topologically well defined, depending on
the specific kind of interface. It is a well defined
topological invariant across a domain wall where the gap
changes sign, but it is not for a BLG-vacuum interface. Such
a marginal topological character of a single valley is
likely to be not only characteristic of BLG, but is a more
general property common to many other multi-valley systems.

\section{Absence of valley-specific bulk-edge correspondence}

\begin{figure}
\begin{center}
 \includegraphics[width=0.43\textwidth]{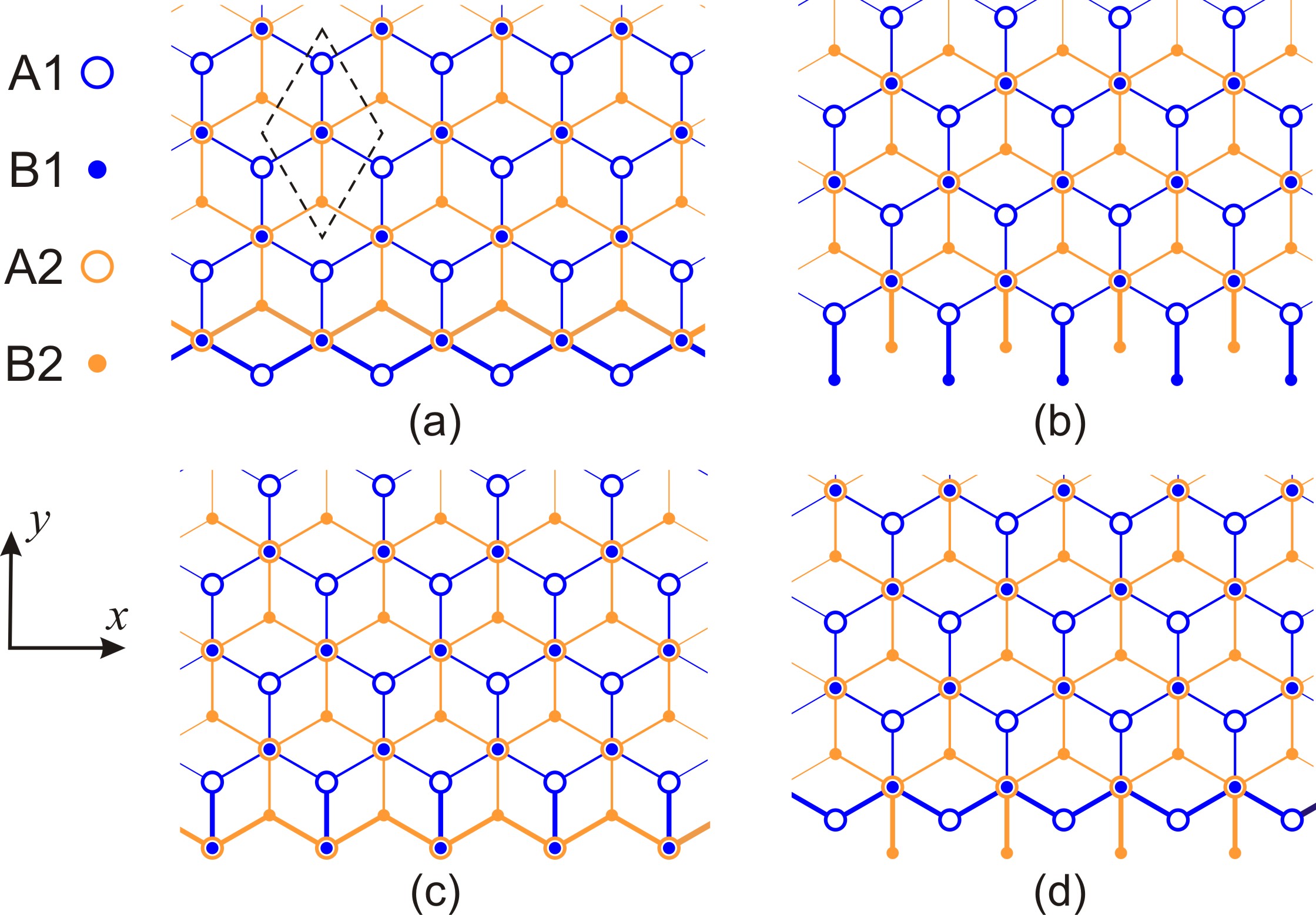}
\end{center}
\caption{Illustration of bilayer graphene edges (highlighted
by thick lines) terminating all in the zigzag direction but
with different sublattices. The number of subgap edge modes
is 1 for case (a) and (b), 2 for case (c) and 0 for case
(d). A unit cell of the lattice is also shown in (a) in
broken lines.} \label{fig:lattice}
\end{figure}

We start by briefly reviewing the known properties of the
band structure of bilayer graphene, emphasizing the specific
aspects that will be relevant later. In BLG a gap between
valence and conductance bands can be opened controllably by
applying a perpendicular electric field, while maintaining
the Fermi energy in the middle of the
gap\cite{castro_biased_2007,
oostinga_gate-induced_2008,zhang_direct_2009} (the sign of
the gap is determined by the orientation of the field). The
continuum low-energy, long-wavelength limit Hamiltonian for
each individual valley can be obtained from the
tight-binding description of BLG with Bernal stacking
\cite{mccann_landau-level_2006}. With the four inequivalent
atoms in a unit cell labeled by A(B)1(2) (see Fig.
\ref{fig:lattice}(a)), the dimensionless Hamiltonian matrix
for the wave function $\Psi = (\chi_{B1}, \chi_{A2},
\varphi_{B2}, \varphi_{A1})^T$ reads \cite{footnote2}:
\begin{align}\label{eq:ham4}
&H_{\tau} = \left(\begin{array}{cc}
H_1 & K_{\tau}  \\
K_{\tau} & H_0
\end{array}\right), \\
&H_1 = \sigma_x + \Delta\sigma_z,\; H_0 = -\Delta\sigma_z \nonumber\\
&K_{\tau} = K_{\tau}^{\dagger} = \tau k_x\sigma_x +
k_y\sigma_y.\nonumber
\end{align}
Here, $\bm{k}= (k_x,k_y)$ is the (dimensionless) wave
vector, the valley index $\tau = \pm 1$, and $2|\Delta|$
($|\Delta| \ll 1$) defines the size of the bulk gap. The
excitation spectrum is $E_{h,l} \simeq \pm
\sqrt{\Delta^2+(k^2+\epsilon^2_{h,l})^2}$, with
$\epsilon_{h} = 1$ and $\epsilon_{l} = 0$. This implies a
splitting of an order of magnitude $1$ between the high
energy bands ($h$)  and direct gap $2|\Delta|$ between the
low energy bands ($l$).

The original four-component Hamiltonian \eqref{eq:ham4} can
be inconvenient to work with, especially when one needs to
find solutions with specific boundary conditions. Therefore,
bearing in mind that the energy range of our interests would
only involve the two low energy bands, we reduce
\eqref{eq:ham4} to a two-component Hamiltonian by rewriting
the Schr\"{o}dinger equation $H_{\tau}\Psi=E\Psi$ in the
following form:
\begin{eqnarray}
&\chi = -(E + \sigma_x + \Delta\sigma_z)K_{\tau} \varphi,& \label{eq:chi} \\
&-[K_{\tau}\sigma_x K_{\tau} + \Delta(1-k^2)\sigma_z]
\varphi = E(1+k^2) \varphi,& \label{eq:phi}
\end{eqnarray}
where $\varphi= (\varphi_{B2}, \varphi_{A1})^T$ and $\chi = (\chi_{B1},
\chi_{A2})^T$.
As we are interested in the subgap edge state solutions with
$|E| < |\Delta| \ll 1$, only terms up to linear order in $E$
and $\Delta$ are kept. If we further neglect $O(\Delta k^2)$
and $O(E k^2)$ terms, we obtain
\begin{equation}\label{eq:ham2}
-\left(\begin{array}{cc}
\Delta & (\tau k_x - ik_y)^2  \\
(\tau k_x + ik_y)^2 & -\Delta
\end{array}\right) \varphi = E \varphi,
\end{equation}
which is the reduced Hamiltonian, describing the low energy
bands to the lowest order in $\Delta$ and $k$. This
reduction procedure is essentially identical to that used by
McCann and Fal'ko \cite{mccann_landau-level_2006}. Here we
emphasize two important aspects: first, the nontrivial
topological properties of the original four-component model
are fully inherited by the reduced two-component model --
this will be shown in Section \ref{ssec:bulk}; second, the
relation between the low-energy ($\varphi$) and high energy
($\chi$) components of the wavefunction [Eq. \eqref{eq:chi}]
is essential to impose the correct boundary conditions at
edges -- this will be shown in Section \ref{ssec:edge}.

\subsection{Quantized Hall conductivity of individual valleys} \label{ssec:bulk}

Starting from the effective Hamiltonians given previously
and following the logic explained in the introduction, we
examine the nontrivial topological properties of a single
valley of gapped BLG by calculating the corresponding
(single-valley) contribution to the Hall conductivity
$\sigma_H^{\tau}$. We consider the case when the Fermi
energy is lying in the middle of the bulk energy gap, so
that we are dealing with one completely full and one
completely empty bulk bands of \eqref{eq:ham2}. The
valley-specific Hall conductivity $\sigma_H^{\tau}$ (in units of $e^2/h$)
corresponds, through Kubo formula, to the quantity
$N_{\tau}$ defined previously. Here we present calculations
for both the original four-band model \eqref{eq:ham4} and
the reduced two-band model \eqref{eq:ham2}, and show that
these calculations yield consistent results. This
consistency implies that the nontrivial momentum topology of
the original four-band model is fully inherited by the
two-band model, which lends itself more easily to a
geometrical interpretation of our results.

The Hall conductivity (in units of $e^2/h$) associated to a
fully-gapped model is generically given
by\cite{volovik_analog_1988, yakovenko_chern-simons_1990}
\begin{align}\label{eq:tc4}
N = \frac{e_{\mu\nu\lambda}}{24\pi^2}\mbox{Tr}\int d^2k dk_0
G\partial_{k_\mu}G^{-1} G\partial_{k_\nu}G^{-1}
G\partial_{k_\lambda}G^{-1},
\end{align}
where the non-singular propagator $G(k_0,\bm{k})$ is defined
as $G \equiv (ik_0 - H)^{-1}$. For the four-component
single-valley Hamiltonian of BLG \eqref{eq:ham4}, we find
$N_{\tau} = \tau\,\mbox{sign}(\Delta)$, which is equal and
magnitude but opposite in sign ($\pm 1)$ for two valleys;
therefore, the Chern number defined over the entire
Brillouin zone vanishes. This result is expected for BLG,
which is strictly speaking topologically trivial, despite
the fact that the individual valleys possess nontrivial
$N_\tau$.

For the reduced two-component model given by Eq.
\eqref{eq:ham2}, we can write the Hamiltonian as
$H_{\tau}^{(2\times 2)} =- {\bf g}_{\tau}(\bm{k}) \cdot {\bm
\sigma}$, where ${\bf g}_{\tau}(\bm{k}) = (k_x^2-k_y^2,
2\tau k_x k_y, \Delta)$. In this case, using the simplified
form of the above formula for two-band
models\cite{volovik_universe_2003}, given by
\eqref{eq:intro2}, we find $N_{\tau} =
\tau\,\mbox{sign}(\Delta)$, in agreement with the
calculation based on the full four-band model. This
agreement indicates that the nontrivial topological
properties of the individual valleys are due solely to the
low energy bands.

It is worth noting that the integral (divided by 2) in Eq.
\eqref{eq:intro2} can be also identified as the Berry phase
acquired by an electron adiabatically transferred along the
path enclosing the area of the integral
\cite{xiao_berry_2010}. In this sense, the quantized Hall
conductivity is equivalent to the quantized Berry phase (in
units of $2\pi$) associated with the single-valley
Hamiltonian \cite{novoselov_unconventional_2006}.

\subsection{Boundary conditions and zero-energy edge states} \label{ssec:edge}

We now proceed to calculate the exact solution of the edge
states in gapped BLG, for different edge structures that do
not mix the valleys. We will only consider one of
the two valleys, with $\tau=+1$, and the results for the
other valley can be inferred through a symmetry
transformation similar to time reversal.

\begin{figure}
\begin{center}
  \includegraphics[width=0.42\textwidth]{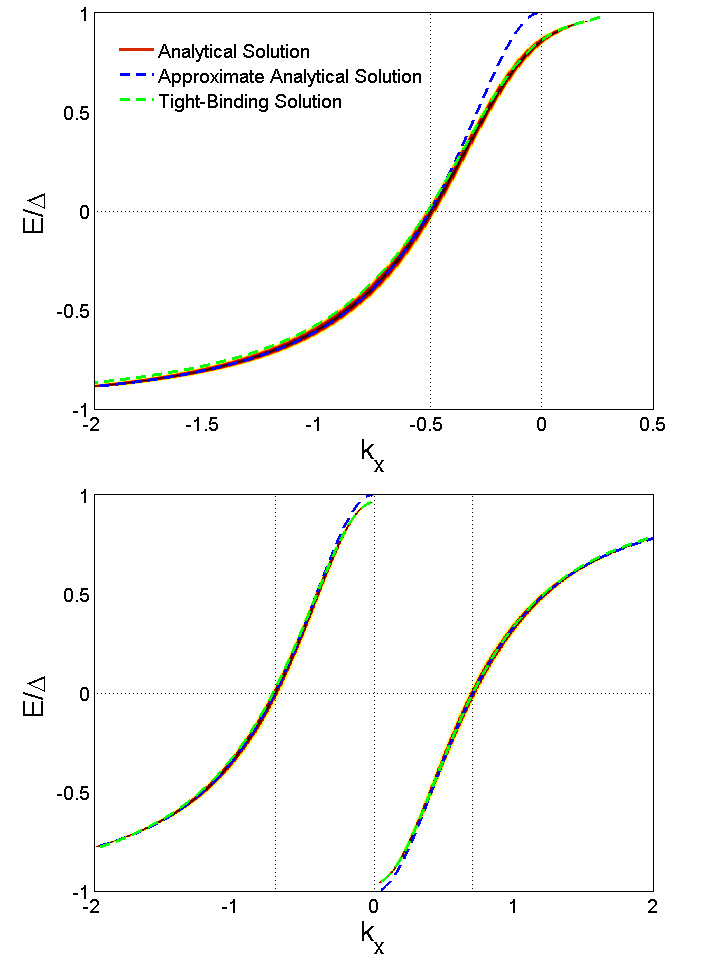}
\end{center}
\caption{Comparison of the dispersion relations for the
subgap edge modes obtained by solving the continuum model
and exact diagonalization of the tight-binding model, with
$\Delta=0.15$. The upper panel case corresponds to the edge
shown in Fig. \ref{fig:lattice} (a), and contains only one
branch of gapless edge modes; the lower panel case
corresponds to Fig. \ref{fig:lattice} (c), and contains two
branches of gapless edge modes.} \label{fig:solutions}
\end{figure}

The generic constraint for boundary conditions in the
present model is imposed by vanishing probability current
across the boundary. This constraint allows for a variety of
boundary conditions which are physically valid (see Appendix
\ref{app:bc}). In order to base our investigation on
concrete physical examples, we illustrate the boundary
conditions that will be discussed in the following in the
tight-binding picture of BLG. Fig. \ref{fig:lattice} shows
the edge structure for four cases, where the BLG lattice
terminates at different sublattices, either A or B, in each
of the two layers. Indeed, each of these combinations
corresponds to a distinct boundary condition imposed in the
continuum model, similar in spirit to what has been
discussed by Brey and Fertig \cite{brey_electronic_2006} for
single layer graphene. A significant difference is that in
BLG the boundary conditions in general involve both low
($\varphi$) and high ($\chi$) energy components of the
wavefunction. Still, thanks to Eq. \eqref{eq:chi}, the
latter can be written in terms of the low energy components
and their derivatives, which is why a description in terms
of the low energy components only is possible. For the
structure in Fig. \ref{fig:lattice}(a) where the BLG lattice
terminates at the edge with sublattices A1 and A2, this
corresponds to imposing $\varphi_{B2}(y=0) = \chi_{B1}(y=0)
=0$  (the BLG is infinitely long in the $x$ direction and
semi-infinite in the $+y$ direction, i.e. the edge is
located at $y=0$). The general solution of \eqref{eq:phi}
for $|E|<|\Delta|$ is
\begin{align} \label{eq:gen_sol2}
& \varphi = \sum\limits_{s=\pm} c_s \left(\begin{array}{c}
1  \\
-\frac{(\Delta + E) - (\Delta - E)(k_x^2 - \kappa_s^2)}{(k_x
+ \kappa_s)^2}
\end{array}\right)
e^{ik_x x - \kappa_s y},
\end{align}
with
\begin{align} \label{eq:kappa2}
\kappa_{\pm}^2 = k_x^2 - (\Delta^2 + E^2) \pm
i\sqrt{\Delta^2-E^2},\; \Re(\kappa_{\pm})>0,
\end{align}
and $c_s$ ($s=\pm$) are coefficients that need to be determined. Using
\eqref{eq:chi}, the boundary conditions lead to two coupled
equations for $c_s$ with the secular equation for the existence
of nontrivial subgap solutions
\begin{align}
(k_x+\kappa_+)(k_x+\kappa_-) = (\Delta + E)^2.
\label{eq:BCa}
\end{align}
This equation gives the dispersion relation $E(k_x)$ for one
sub-gap edge mode. When $|E|,|\Delta|\ll k_x^2$ (as is true
for the $E=0$ mode) Eq. \eqref{eq:BCa} can be simplified
(since $\kappa_{\pm} \simeq |k_x| \pm
i(\sqrt{\Delta^2-E^2})/{2|k_x|}$) to:
\begin{align}
\frac{E}{\Delta} = \frac{1-4k_x^2}{1+4k_x^2}, \quad k_x<0.
\label{eq:BCa_ap}
\end{align}
In Fig. \ref{fig:solutions}(a) the analytical solution is
compared with the full tight-binding solution, from which we
see that the continuum model represents an excellent
approximation. We conclude that for the edge just
considered, one zero-mode exists (per valley and spin)
\cite{castro_localized_2008}, i.e. the result expected if
the bulk-edge correspondence for single valley holds.

We now consider the edge shown in Fig. \ref{fig:lattice}(c).
The corresponding boundary conditions are $\varphi_{B2}(y=0)
= \varphi_{A1}(y=0) = 0$, which lead to the new dispersion
relation for the subgap edge modes, given by
\begin{align}
\frac{2k_x(k_x+\kappa_+)(k_x+\kappa_-)}{2k_x+\kappa_++\kappa_-}
= \frac{\Delta + E}{\Delta - E}. \label{eq:BCc}
\end{align}
This equation has two solutions related by transformation
$k_x\rightarrow -k_x$ and $E\rightarrow -E$, which comes
from the symmetry of  the wave equation \eqref{eq:phi} under
$\{k_x, E, \varphi\}\rightarrow \{-k_x, -E,
i\sigma_y\varphi\}$. It is this symmetry, which is also
preserved by the boundary conditions in the present case
(whereas it is broken by the boundary conditions considered
in the previous example), that guarantees the existence of
\textit{two} subgap edge modes. The dispersion relation
given by \eqref{eq:BCc} can be further simplified when
$|E|,|\Delta|\ll k_x^2$, to
\begin{align}
\frac{E}{\Delta} =
-\mbox{sgn}(k_x)\frac{1-2k_x^2}{1+2k_x^2}, \label{eq:BCc_ap}
\end{align}
which explicitly shows a pair of gapless edge modes
propagating in the same direction, plotted in Fig.
\ref{fig:solutions}(b) (again, in excellent agreement with
the tight-binding solution).  Without presenting the details
of the solutions for the other cases, we only state that the
edge shown in Fig. \ref{fig:lattice}(b) ($\varphi_{A1}(y=0)
= \chi_{A2}(y=0) =0$) leads to a single branch of subgap
edge modes, similar to the first example above, and the one
shown in Fig. \ref{fig:lattice}(d) ($\chi_{B1}(y=0) =
\chi_{A2}(y=0) = 0$) yields no subgap modes. We therefore
have to conclude that the number of zero-modes at the
BLG-vacuum interface depends on the specific edge
considered, i.e., the bulk-edge correspondence relating the
Hall conductivity and gapless edge modes does not hold in
general for individual valleys \cite{footnote3}.

\section{Discussion}

Finding that bulk-edge correspondence does not hold for an
individual valley, even though the contribution that each
valley gives to the Hall conductivity is integer (in units
of $e^2/h$) may seem surprising. Specifically, one may
wonder why the argument based on Laughlin's gedanken
experiment \cite{laughlin_quantized_1981} -- which is
normally invoked to justify the existence of edge states in
integer quantum Hall systems -- does not apply to individual
valleys in gapped BLG. The answer to this question has to do
with the fact that Laughlin's gedanken experiment considers
the adiabatic flow of electric charge, which is a conserved
quantity (in the context of Laughlin's argument conservation
of charge is so obvious that this assumption is not normally
emphasized). Contrary to the charge, the valley quantum
number is not in general conserved. In particular, it is not
conserved even in the presence of ideal edges if one
considers the process -- the adiabatic insertion of flux,
which is  the basis of Laughlin's gedanken experiment. As we
outline below, the non-conservation of the valley quantum
number allows for the possibility of a quantized valley Hall
conductivity in the absence of zero-energy edge modes.

Upon varying the magnetic flux $\phi$ in the usual
cylindrical geometry involved in Laughlin's argument, the
allowed $k_x$ values of momentum along the edge  flow
according to $k_x\to k_x + \phi/L$, where $L$ is the
circumference of the cylinder.  Upon insertion of flux $\phi
= 2 \pi$, the set of allowed $k_x$ values returns to the
original one. From the generic form of the electron wave
function \eqref{eq:gen_sol2} it follows that the
wave-function weight is redistributed in the direction
perpendicular to the edge in the process of flux insertion.
The redistribution occurs in the opposite directions for the
states near two valleys, with states in one valley flowing
towards the edge, and states in the other valley flowing
away from the edge. This is the microscopic mechanism for
the transverse valley current flow implied by the non-zero
value of the valley Hall conductivity, $\sigma_H^v = N_\tau
- N_{-\tau}$. The fact that the valley current flows
perpendicular to the edge, according to continuity equation
would imply accumulation of the valley density at the edge.
This apparent valley charge accumulation can be accommodated
in two ways. The first is by creating valley charge
imbalance at the edge, in direct analogy to quantum Hall and
quantum spin Hall effect with conserved spin. Clearly, for
that to occur, the gapless edge modes have to exist.
Alternatively, the valley current influx at the edge can be
compensated by the adiabatic transfer of the electrons
between the valleys induced by the flux insertion, which
converts $k_x\to k_x + 2\pi/L$ for states in the {\em
entire} 1D Brillouin zone of the cylinder.  This is possible
due to the fact that the valleys are indeed connected by the
bands that extend deep below the chemical potential. As a
consequence, the spectral flow cannot be fully described in
terms of the long-wavelength effective Hamiltonian in the
neighborhood of the $K$ and $K'$ points, but it requires
considering the Hamiltonian of the system over the entire
Brillouin zone. This is how the non-conservation of the
valley quantum number can account for the seemingly
paradoxical result that non-zero $\sigma_H^v$ can exist even
when gapless edge modes are completely absent
\cite{Martin_notes_2010}.

\begin{figure}
\begin{center}
  \includegraphics[width=0.37\textwidth]{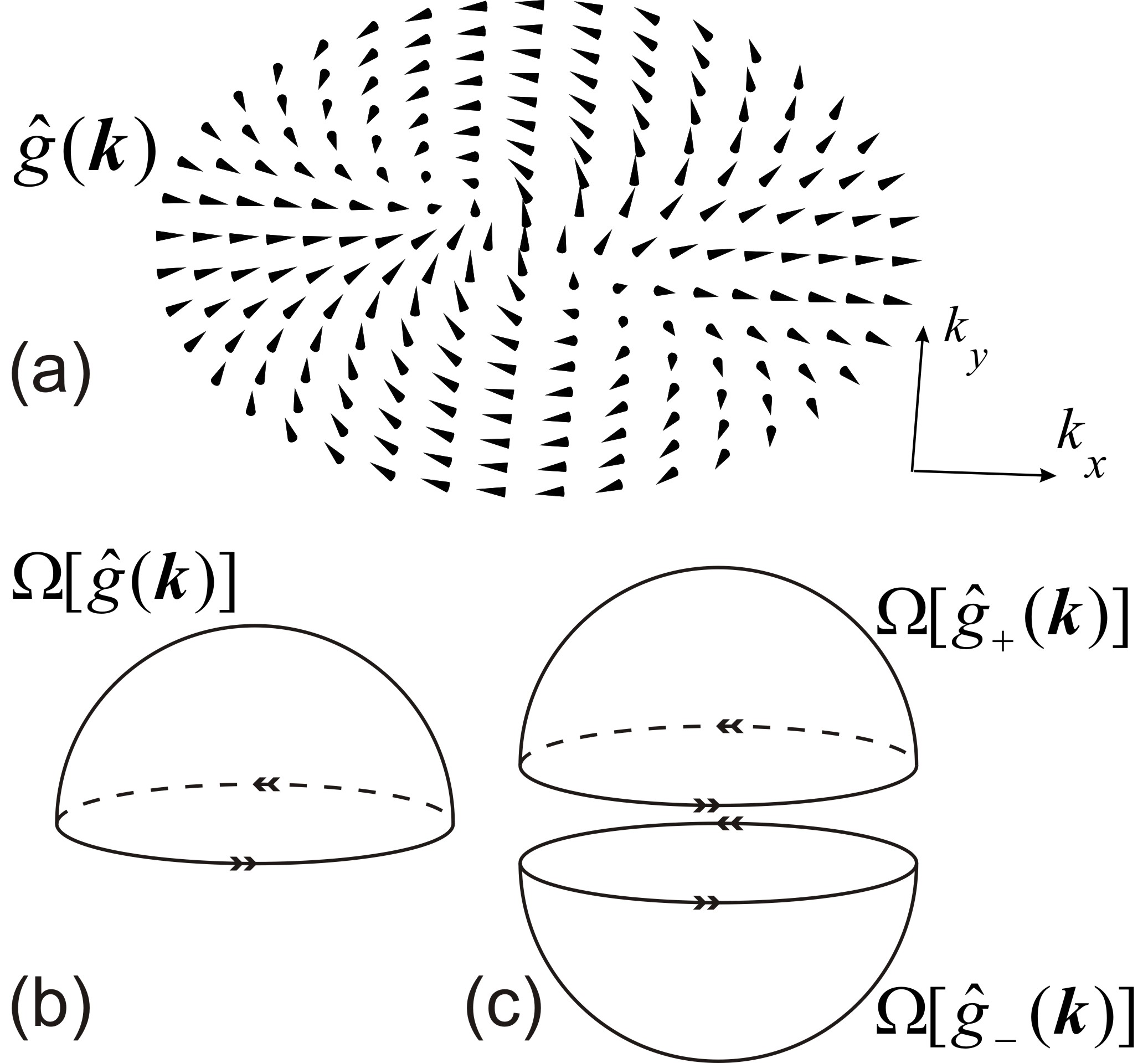}
\end{center}
\caption{Illustration of the mapping $\hat{\bf g}(\bm{k})$
for gapped BLG. Panel (a) shows $\hat{\bf g}(\bm{k})$ for a
single valley as a vector field on the $R^2$ plane of
$\bm{k}$. Panel (b) shows the solid angle $\Omega$ that
$\hat{\bf g}$ covers when $\bm{k}$ runs over the whole
plane, the double arrow signifying the double covering of
the (upper) hemisphere. Panel (c) shows how two marginal
topological mappings can be ``glued" to form a well-defined
topological mapping, which is the case for a BLG-BLG domain
wall with opposite mass signs on the two sides.}
\label{fig:topo}
\end{figure}

Next, let us give a geometrical reason for the the absence
of a valley-specific bulk-edge correspondence at gapped BLG
edges (BLG-vacuum interfaces).  It originates from the fact
that the quantity $N_{\tau}$ associated with a given valley
is not a well-defined topological invariant. This statement
may appear to be in conflict with the case of domain wall
interfaces, across which the sign of the mass ($\Delta$) in
BLG changes sign\cite{martin_topological_2008}. In this
latter case, bulk-edge correspondence ``predicts" that
gapless states should be present, whose number corresponds
to $N_{\tau}^l-N_{\tau}^r$. When the gap $\Delta$ changes
sign across the domain wall, $|N_{\tau}^l-N_{\tau}^r|=2$ and
indeed two gapless states are found at the interface. The
conflict is only apparent, because
$N_{\tau}^l-N_{\tau}^r$ does correspond to a well-defined
topological invariant, even though $N_{\tau}$ does not.

To understand the difference between $N_{\tau}$ and
$N_{\tau}^l-N_{\tau}^r$ we look at the integrand function
that is used to calculate these quantities, which, from Eq.
\eqref{eq:intro2}, is given by ${\mathcal B}_k =
\frac{1}{4\pi} \hat{\bf g}\cdot [\partial_{k_x} \hat{\bf
g}\times\partial_{k_y} \hat{\bf g}]$. This expression
corresponds to the Jacobian of the transformation from the
region $R^2$ of the momentum space in the vicinity of a
given valley point, to the unit sphere $S^2$ on which $\bf
\hat g$ resides. For BLG, at large $k$ ($k^2 \gg |\Delta|$),
the vector $\hat{\bf g}$ lies on the equator, and therefore
the integral of ${\mathcal B}_k$ does not represent a
topological invariant since the mapping between non-compact
$R^2$ and compact $S^2$ is topologically trivial. Indeed, as
shown in Fig. \ref{fig:topo} (a) and (b), the vector
$\hat{\bf g}$ wraps around half of the unit sphere twice,
but does not cover the entire sphere. That is despite the
fact that the integral rapidly converges for $k^2 \gg
|\Delta|$ and is equal to 1 (for a given valley and sign of
the gap $\Delta$; it is $-1$ for the opposite valley or for
opposite sign of $\Delta$). For massive Dirac fermions such
non-compact behavior has been discussed by Volovik, and is
known as {\em marginal} \cite{volovik_universe_2003}. In the
same sense, the massive chiral fermions in BLG  are marginal
as well. The marginality implies that a small perturbation
in the Hamiltonian (e.g. momentum-dependent mass term
$\Delta(k^2)$) or boundary conditions can have a big effect
on presence or absence of the gapless modes (e.g.
\cite{yao_edge_2009}).

Now, for a domain wall such that $\Delta(x<0) <0$ and
$\Delta(x>0) >0$, despite the marginal character of
$N_{\tau}$ on the two sides of the domain wall, their
difference is a well-defined topological invariant.  That is
because at $k\rightarrow \infty$, the $\Delta$ in the
Hamiltonians becomes irrelevant and the textures of the
$\bf\hat g$ vectors for the two insulators seamlessly
connect on the equator of $S^2$, thereby compactifying the
momentum space {\em within the same valley}. The connection
of the textures is shown schematically in Fig.
\ref{fig:topo} (c). Similar considerations are not only
valid for BLG, but can also be extended to individual
valleys in (gapped) single layer graphene
\cite{semenoff_domain_2008}, as well as to interfaces
between gapped single layer graphene and Kane-Mele
topological insulators with conserved spin
\cite{kane_quantum_2005}. In these cases, the differences in
the single-valley $N_{\tau}$ across the interface is a well
defined topological invariant, corresponding to number of
gapless states present (or more precisely, the difference
between the numbers of left and right moving gapless modes
in a given valley). Mathematically, this is a consequence of
the index theorem discussed in this type of contexts by
Volovik (see Section 22.1.4 in
\cite{volovik_universe_2003}).

\section{Conclusion}

Our work shows that the long wavelength Hamiltonians that
describe the low-energy electronic states in individual
valleys of gapped BLG possess non-trivial topological
features. In contrast to integer quantum Hall systems and
quantum spin Hall systems (with conserved spin), these
non-trivial valley-specific properties cannot be described
by a topological invariant. Rather, they require the
analysis of the specific interfaces (in the present case the
BLG-vacuum interface and the domain wall) in order to
establish whether a  topological
invariant that determines the number of gapless edge modes
can be defined. This is the characteristic signature of
marginal topological insulators.

It is certainly the case that the low-energy edge states
that can be predicted through these topological
considerations are not robust against short-range disorder
that couples the valleys (even though they are stable
against long range disorder that scatters electrons within
the same valley). In the presence of inter-valley
scattering, the gapless modes associated to individual
valleys will couple, leading to the opening of a transport
gap. Nevertheless, this does not mean that the low-energy
states at interfaces and domain walls are experimentally
irrelevant. In fact, for sufficiently pure materials when
the Fermi energy is in the gap, the states originating from
the localization of the gapless modes can provide a dominant
path for transport in the insulating state, because no other
states are available deep in the bulk gap of the material
\cite{li_topological_2010}.

Finally, even though all calculations presented here have
been performed specifically for gapped BLG, our arguments do
not rely on the particular form of the valley Hamiltonian,
or the number of valleys. Consequently, similar treatment
based on other effective valley Hamiltonians should be
applicable to many other materials and interfaces, and may
provide a convenient tool for determining the interfacial
electronic properties.

\begin{acknowledgments}
This work has been supported by the Swiss National Science
Foundation (projects 200020-121807, 200021-121569), by the
Swiss Center of Excellence MaNEP, and the European Network
NanoCTM. The work of IM was carried out under the auspices
of the National Nuclear Security Administration of the U.S.
Department of Energy at Los Alamos National Laboratory under
Contract No. DE-AC52-06NA25396 and supported by the
LANL/LDRD Program.
\end{acknowledgments}

\section{Appendix: Boundary conditions}\label{app:bc}

In this appendix we derive the most general boundary conditions
from the continuum model given by \eqref{eq:ham4} or
\eqref{eq:ham2}, by imposing current conservation
through a hard-wall boundary. We show that the
boundary conditions associated to the edge structures considered
in the main text are specific realization of the general case.

Let us start from the original four-component model
\eqref{eq:ham4}, for which the current operator is given by
\begin{align}
\bm{j} = \left(\begin{array}{cc}
0 & \bm{\sigma} \\
\bm{\sigma} & 0
\end{array}\right), \label{eq:cur_op}
\end{align}
where $\bm{\sigma}\equiv(\sigma_x,\sigma_y)$ and we have let
$\hbar=1$. Without losing generality, we consider a boundary
at $y=0$ with unit normal $\bm{n}_B=-\hat{y}$; then, the
current conservation implies
\begin{align}
\Psi^\dagger(\bm{j}\cdot\bm{n}_B)\Psi\bigr|_{y=0} =
i(\chi_{B1}^*\varphi_{A1}-\chi_{A2}^*\varphi_{B2})\bigr|_{y=0}
+ c.c. = 0. \label{eq:cur_cons}
\end{align}
We immediately see that the above condition can be naturally
satisfied by letting in each layer the wave amplitude on either A
or B component vanish at $y=0$. This leads to four choices
of combinations each finding an exact correspondence to the
cases listed in Fig. \ref{fig:lattice}. Indeed, the same
amount of freedom in ``choosing" boundary conditions is also
inherited by the reduced model \eqref{eq:ham2}, for which,
by the same token, the vanishing current across the boundary
requires
\begin{align}
(\varphi_{A1}^*\partial_+\varphi_{B2}-\varphi_{B2}^*\partial_-\varphi_{A1})\bigr|_{y=0}
+ c.c. = 0, \label{eq:cur_cons2}
\end{align}
where $\partial_\pm\equiv\partial_x\pm i\partial_y$.
Noticing that the leading order approximation of Eq.
\eqref{eq:chi} yields
\begin{align}
\chi_{B1} = i\partial_+\varphi_{B2},\quad \chi_{A2} =
i\partial_-\varphi_{A1}, \label{eq:chi2}
\end{align}
Eq. \eqref{eq:cur_cons2} is nothing but Eq.
\eqref{eq:cur_cons}. In other words, the reduced model
shares the same choices of boundary conditions as the
original one, and the edge modes it allows for may vary
accordingly.

These derivations indicate that for a generic multi-component
Hamiltonian, considerable freedom exist in selecting boundary conditions
associated to the wave-equation, reflecting different possible structures of the edges.
To demonstrate the dependence of the gapless edge modes on boundary conditions, which serves the purpose of
the current work, however, the four cases presented in the main text are already sufficient.


\begin{thebibliography}{30}
\bibitem[{foo({\natexlab{b}})}]{footnote2}
\bibinfo{note}{To make the Hamiltonian and all its parameters dimensionless, we
  scale energy with $t_{\perp}$, and wave vector with $1/l_0 \equiv
  2t_{\perp}/3ta$, where $t_{\perp}$ is the inter-layer hopping energy, and $a$
  and $t$ are the intra-layer nearest neighbor distance and hopping energy.}
\bibitem[{foo({\natexlab{c}})}]{footnote3}
\bibinfo{note}{Similarly, there is no correspondence between the Hall
  conductivity and the edge modes in gapped single layer graphene. In fact,
  there is a close connection between the edge states in a graphene single
  layer and its bilayer, as the latter can be obtained by considering two
  decoupled single layers, upon turning on $t_\perp$.}
\end{thebibliography}

\end{document}